\documentclass[aps,prd,twocolumn,amsmath,superscriptaddress,amssymb,showpacs,floatfix,nofootinbib,longbibliography]{revtex4-1}
\pdfoutput=1
\usepackage{graphicx}
\usepackage{bm}
\usepackage{times}
\usepackage{slashed}
\usepackage[usenames,dvipsnames]{xcolor}
\usepackage{aas_macros}
\usepackage{slashed}
\usepackage{lipsum}
\usepackage{subfigure}
\usepackage{multirow}
\usepackage{amsmath}
\usepackage{array} 
\usepackage{varwidth} 
\usepackage[normalem]{ulem}

\usepackage[colorlinks  = true,
            linkcolor   = NavyBlue,
            urlcolor    = NavyBlue,
            citecolor   = NavyBlue,
            anchorcolor = NavyBlue]{hyperref}

\newcommand{\be}{\begin{equation}}
\newcommand{\ee}{\end{equation}}
\newcommand{\bea}{\begin{eqnarray}}
\newcommand{\eea}{\end{eqnarray}}

\let\vec\bm

\newcommand{\diff}{\ensuremath{\mathrm{d}}}

\newcommand{\vx}{\vec x}

\newcommand{\ac}{a_\mathrm{coll}}
\newcommand{\rc}{r_\mathrm{core}}
\newcommand{\Rc}{r_\mathrm{cusp}}

\newcommand{\fmax}{f_\mathrm{max}}
\newcommand{\rhoc}{\bar\rho_0}
\newcommand{\sigmav}{\langle\sigma v\rangle}
\newcommand{\Td}{T_\mathrm{d}}

\newcommand{\kfs}{k_\mathrm{fs}}

\newcommand{\rhoeff}{\rho_\mathrm{eff}}
\newcommand{\MDM}{M}
\newcommand{\gfac}{f_\mathrm{glob}}
\newcommand{\lfac}{f_\mathrm{gal}}

\begin{document}

\title{Limits on dark matter annihilation in prompt cusps from the isotropic gamma-ray background}
\author{M. Sten Delos}
\email{mdelos@carnegiescience.edu, orcid.org/0000-0003-3808-5321}
\affiliation{The Observatories of the Carnegie Institution for Science, 813 Santa Barbara Street, Pasadena, CA 91101, USA}
\affiliation{Max Planck Institute for Astrophysics, Karl-Schwarzschild-Stra{\ss}e 1, 85748 Garching, Germany}

\author{Michael Korsmeier}
\email{michael.korsmeier@fysik.su.se, orcid.org/0000-0003-3478-888X}
\affiliation{Stockholm University and The Oskar Klein Centre for Cosmoparticle Physics,  Alba Nova, 10691 Stockholm, Sweden}

\author{Axel Widmark}
\email{axel.widmark@fysik.su.se, orcid.org/0000-0001-5686-3743}
\affiliation{Stockholm University and The Oskar Klein Centre for Cosmoparticle Physics,  Alba Nova, 10691 Stockholm, Sweden}

\author{Carlos Blanco}
\email{carlosblanco2718@princeton.edu, orcid.org/0000-0001-8971-834X}
\affiliation{Stockholm University and The Oskar Klein Centre for Cosmoparticle Physics,  Alba Nova, 10691 Stockholm, Sweden}
\affiliation{Princeton University, Department of Physics, Princeton, NJ 08544}

\author{Tim Linden}
\email{linden@fysik.su.se, orcid.org/0000-0001-9888-0971}
\affiliation{Stockholm University and The Oskar Klein Centre for Cosmoparticle Physics,  Alba Nova, 10691 Stockholm, Sweden}

\author{Simon D. M. White}
\email{swhite@mpa-garching.mpg.de, orcid.org/0000-0002-1061-6154}
\affiliation{Max Planck Institute for Astrophysics, Karl-Schwarzschild-Stra{\ss}e 1, 85748 Garching, Germany}

\begin{abstract}
Recent studies indicate that thermally produced dark matter will form highly concentrated, low-mass cusps in the early universe that often survive until the present. While these cusps contain a small fraction of the dark matter, their high density significantly increases the expected $\gamma$-ray flux from dark matter annihilation, particularly in searches of large angular regions. We utilize 14~years of Fermi-LAT data to set strong constraints on dark matter annihilation through a detailed study of the isotropic $\gamma$-ray background, excluding with 95\% confidence dark matter annihilation to $b\bar{b}$ final states for dark matter masses below 120~GeV. 
\end{abstract}

\maketitle

\section{Introduction}

While the particle properties of dark matter are not understood, its gravitational properties are strongly constrained by observations across a wide range of distance scales from the cosmic microwave background~\cite{Planck:2018vyg} to the smallest dwarf galaxies~\cite{2019ARA&A..57..375S}. These constraints require only that dark matter has a small scattering cross section and small thermal motions in the low-redshift universe. Under these conditions, both theoretical arguments and simulations imply that dark matter structures grow hierarchically from the ``bottom up'', producing quasi-equilibrium dark matter halos with density profiles that diverge approximately as $r^{-1}$ in their inner region~\cite{Navarro:1996gj}.

Recently, several studies have provided an important extension to theoretical expectations for the small-scale density distributions of gravitationally interacting dark matter. They have shown that, so long as the dark matter is initially smooth on the smallest scales, a pronounced cusp forms at the moment of collapse of each peak in the initial mass density field~\cite{Delos:2019mxl,Delos:2022yhn,Ondaro-Mallea:2023qat}. Unlike the dark matter profiles produced subsequently by accretion and mergers, these ``prompt cusps'' have strongly peaked dark matter densities which rise as $\rho\propto r^{-1.5}$ outside a very small central core set by the initial phase-space density of the dark matter~\citep{2010ApJ...723L.195I,2013JCAP...04..009A,2014ApJ...788...27I,2015MNRAS.450.2172P,2017MNRAS.471.4687A,2018MNRAS.473.4339O,2018PhRvD..97d1303D,2018PhRvD..98f3527D,Delos:2019mxl,2020MNRAS.492.3662I,2021A&A...647A..66C,2022MNRAS.517L..46W,Delos:2022yhn,DelPopolo:2023tcp,Ondaro-Mallea:2023qat}. Contrary to previous expectations, these cusps can survive until the present era unless they are disrupted by tidal encounters with individual stars~\citep{Stucker:2023rjr}, a process that is only important for the small fraction of cusps that reside near the centers of galaxies \cite{Stucker:2023rjr,Delos:2022bhp}. 

\begin{figure}[tbp]
\centering
\includegraphics[width=\columnwidth]{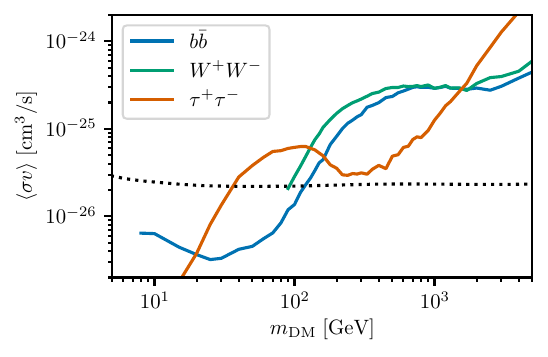}
\vspace{-0.8cm}
\caption{Upper limits on the WIMP annihilation cross-section, accounting for prompt cusps that form at early times and persist throughout the bulk of both the Milky Way and extragalactic halos. The dotted line shows the expected thermal WIMP annihilation cross-section~\cite{Steigman:2012nb}. For dark matter with the thermal cross section, constraints from the isotropic $\gamma$-ray background obtained from our analysis exclude at 95\% confidence masses below 120~GeV annihilating to $b\bar{b}$ and masses around 90~GeV annihilating to W$^+$W$^-$. Constraints lie near the thermal cross-section for $\tau^+\tau^-$.}
\vspace{-0.5cm}
\label{fig:constraints}
\end{figure}

While prompt cusps only provide $\mathcal{O}(1\%)$ of the total dark matter content and are too low-mass and diffuse to produce measurable dynamical or gravitational lensing effects in cold dark matter scenarios, their extremely high densities greatly enhance the dark matter annihilation signal in models where such interactions are possible. Within the paradigm of weakly interacting massive particles (WIMPs)~\cite{Roszkowski:2017nbc}, prompt cusps dominate the total dark matter annihilation signal from all but the densest regions~\cite{Delos:2022bhp}, significantly changing the flux and morphology of the dark matter annihilation signal. 

Most importantly for $\gamma$-ray searches, the existence of a large number of prompt cusps implies that on astrophysical scales the dark matter annihilation signal traces the distribution of cusp number density, rather than the square of the (smooth) dark matter halo density. This implies that the highest sensitivity is achieved by searches of large regions containing the bulk of the dark matter, rather than by searches focused on regions of high local dark matter density. Note that individual prompt cusps are unlikely to be resolved \cite{Delos:2023azx}.

In this paper, we make a new estimate of the isotropic $\gamma$-ray background (IGRB) based on 14 years of Fermi-LAT data. Modeling the $\gamma$-ray contribution from blazar and non-blazar source classes, as well as the simultaneous contribution from extragalactic star-forming (SF) activity, we find no evidence for an excess and strongly constrain any contribution from dark matter, as seen in Fig.~\ref{fig:constraints}. Comparing our results to the flux predicted from prompt cusps in the Milky Way halo and extragalactic environments, we place upper limits on the dark matter annihilation cross-section, excluding with 95\% confidence the thermal annihilation cross-section for dark matter particle masses below 120~GeV annihilating to $b\bar{b}$. Notably, this limit is in tension with models where dark matter annihilation produces the Galactic Center $\gamma$-ray excess~\cite{Daylan:2014rsa}.

\section{Prompt cusps in WIMP scenarios}\label{sec:cusps}

Prompt cusps have $\rho=Ar^{-1.5}$ density profiles that extend out to a limiting radius $\Rc$. Each prompt cusp formed from the collapse of a local maximum in the initial linear density field, and Ref.~\cite{Delos:2022yhn} showed that the density normalization $A$ and the radius $\Rc$ for each cusp are directly connected to the properties of its progenitor density peak. Specifically, numerical simulations suggest that
\begin{equation}\label{eq:A}
    A\simeq 24\rhoc \ac^{-3/2} R^{3/2}
\end{equation}
is accurate at the 10 percent level for individual prompt cusps. Here $\rhoc$ is the comoving dark matter density, $\ac$ is the cosmic scale factor when the cusp's progenitor peak collapses (according to the ellipsoidal collapse approximation in Ref.~\cite{Sheth:1999su}), and $R\equiv (\delta/|\nabla^2\delta|)^{1/2}$ is the peak's characteristic comoving size, defined in terms of its height $\delta\equiv(\rho-\bar\rho)/\bar\rho$ and curvature $\nabla^2\delta$.
We adopt $\rhoc=33.1$~M$_\odot$\,kpc$^{-3}$ \cite{Planck:2018vyg}.
Furthermore, the relation
\begin{align}\label{eq:Rc}
    \Rc\simeq 0.11\ac R
\end{align}
is accurate to within a factor of a few for individual prompt cusps, with uncertainty dominated by the unclear operational definition of a prompt cusp's edge. Our results are only logarithmically sensitive to $\Rc$. As Ref.~\cite{Delos:2022yhn} noted, Eqs. (\ref{eq:A}) and~(\ref{eq:Rc}) are demanded by dimensional considerations; only the numerical factors are tuned to match simulations.

Since the annihilation rate within a $\rho\propto r^{-1.5}$ density profile diverges logarithmically at small radii, it is also necessary to quantify the minimum radius at which the profile is valid. As noted by Ref.~\cite{Delos:2022yhn}, thermal motion in the initial conditions imposes a maximum phase-space density $\fmax$ that cannot be exceeded by any gravitationally evolved structure. This leads to the requirement that the $\rho\propto r^{-1.5}$ profile transitions to a finite-density core within the core radius \cite{Delos:2022yhn}
\begin{equation}\label{eq:rc}
    \rc\simeq 0.1 G^{-2/3} \fmax^{-4/9} A^{-2/9},
\end{equation}
where $G$ is the gravitational constant. The density of a cored prompt cusp changes from $\rho=Ar^{-1.5}$ for $r\gg\rc$ to $\rho\simeq A\rc^{-1.5}$ for $r\ll\rc$. Here we adopt the cored cusp description of Ref.~\cite{Stucker:2023rjr}, which is derived from an Ansatz for the phase-space density and for which the volume integral of the squared density (out to $r=\Rc$) is
\begin{equation}\label{eq:J}
    J \equiv \int \rho^2 \diff V = 4\pi A^2 [0.531 + \log(\Rc/\rc)].
\end{equation}
The annihilation rate inside a prompt cusp is proportional to $J$. Note that typically $\Rc/\rc\sim 500$ for WIMP models \cite{Delos:2022bhp}, so the quantity in brackets in Eq.~(\ref{eq:J}) is approximately 7.

Prompt cusps accrete dark matter halos around them, which also contribute to the annihilation signal. However, Ref.~\cite{Delos:2022bhp} noted that the annihilation rate in prompt cusps exceeds that predicted by previous halo and subhalo models (e.g.~\cite{Hiroshima:2018kfv}) by at least an order of magnitude in the cosmological average, which suggests that halos only contribute to annihilation around the 10 percent level. We neglect their contribution.

\subsection{The prompt cusp distribution}

\begin{figure}[tbp]
\centering
\includegraphics[width=\columnwidth]{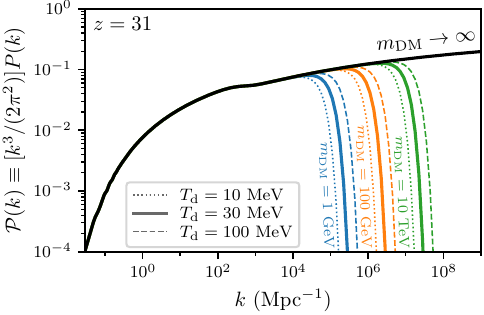}
\caption{Power spectra of dark matter density variations, evaluated in linear theory at $z=31$ and shown in the dimensionless form, $\mathcal{P}(k)\equiv [k^3/(2\pi^2)]P(k)$. The black curve shows the prediction for idealized cold dark matter. Density variations in WIMP models are smoothed by thermal motion in the dark matter, imposing a small-scale cutoff in $\mathcal{P}(k)$ that depends on the particle mass $m_\mathrm{DM}$ and the temperature $\Td$ at which it decoupled from the Standard Model plasma. The colored curves show $\mathcal{P}(k)$ for such scenarios; different colors correspond to different $m_\mathrm{DM}$ while different line styles correspond to different values of $\Td$.}
\label{fig:power}
\end{figure}

We evaluate the abundance of prompt cusps and the distribution of their structural properties as in Ref.~\cite{Delos:2022bhp}.
Due to Eqs. (\ref{eq:A}) and~(\ref{eq:Rc}), these follow directly from the properties of the peaks in the initial linear density field, and thus can be calculated explicitly from the linear power spectrum of dark matter density fluctuations \citep{Bardeen:1985tr}.
Adopting Planck 2018 cosmological parameters \cite{Planck:2018vyg}, we use the \textsc{CLASS} code \cite{2011JCAP...07..034B} to evaluate the linear dark matter power spectrum at redshift $z=31$. The power spectrum at $k>2\times 10^4$~Mpc$^{-1}$ is then evaluated analytically according to Ref.~\cite{Hu:1995en} and matched in amplitude to the \textsc{CLASS} output. The resulting power spectrum is shown as the black curve in Fig.~\ref{fig:power} and is valid for dark matter models with no initial thermal motion. Such models give rise to density variations on arbitrarily small scale.

However, WIMP dark matter was once in equilibrium with the Standard Model plasma, and the resulting thermal motion smooths out density variations below some characteristic free-streaming scale $\kfs^{-1}$. The smoothing scale in the initial dark matter distribution is of fundamental importance to the distribution of prompt cusps \cite{Delos:2023exh}. We evaluate $\kfs$ via Eq.~(48) of Ref.~\cite{Bertschinger:2006nq} and multiply the idealized power spectrum in Fig.~\ref{fig:power} by the Gaussian function $\exp(-k^2/\kfs^2)$. The colored curves show the resulting power spectra for a sample of WIMP scenarios. Particles of lower mass, $m_\mathrm{DM}$, have higher thermal velocities, so they stream over greater distances, leading to lower values for $\kfs$. The temperature $\Td$ at which the dark matter kinetically decoupled from the Standard Model plasma is also relevant; later decoupling (lower $\Td$) also leads to larger streaming distances and hence lower $\kfs$.

For each dark matter power spectrum, we use the procedure described in Appendix C of Ref.~\cite{Delos:2019mxl} to evaluate the number density $n_\mathrm{peaks}$ of initial density peaks and to generate a Monte Carlo sample of $10^7$ such peaks. These peaks are distributed in $\delta$, $\nabla^2\delta$, and the ellipticity $e$ and prolateness $p$ of the tidal field.
We note that this procedure is exact, aside from the distribution of $e$ and $p$, which is still very nearly exact for the power spectra that we consider \cite{Delos:2022bhp}.
Each peak's collapse time follows as $\ac=[\delta_\mathrm{ec}(e,p)/\delta]^{1/g}a$, where $a=1/32$ is the scale factor at which the power spectrum was evaluated, $\delta_\mathrm{ec}(e,p)$ is the threshold for ellipsoidal collapse given by Ref.~\cite{Sheth:1999su},\footnote{For some $e$ and $p$, the approximation in Ref.~\cite{Sheth:1999su} has no solution. We follow Ref.~\cite{Delos:2022bhp} in neglecting these peaks; they have too low $\delta$ for any cusps that they form to significantly impact the annihilation signal anyway.} and $g\simeq 0.901$ \cite{Hu:1995en} due to the baryon-driven suppression of dark matter density perturbations at small scales (an effect responsible for the feature in Fig.~\ref{fig:power} around $k\sim 10^{2.5}$~Mpc$^{-1}$).
The random sample of peaks is then translated into a sample of prompt cusps using Eqs. (\ref{eq:A}--\ref{eq:rc}), with $\fmax$ determined as in Ref.~\cite{Delos:2022bhp}.

\subsection{Annihilation in prompt cusps}\label{sec:rhoeff}

We are interested the annihilation signal from a mass $\MDM$ of dark matter that is large enough to contain many prompt cusps. The volume integral of the squared density contributed by prompt cusps in this case, per $\MDM$, is
\begin{align}\label{eq:rhoeff_samp0}
    \rhoeff
    &\equiv \frac{1}{\MDM}\int_\text{cusps} \rho^2 \diff V
    = \frac{\sum_{\text{cusps } i} J_i}{\MDM}
    \\\label{eq:rhoeff_samp}
    &= n_\mathrm{peaks}\langle J \rangle_\mathrm{sample}/\rhoc
\end{align}
on average, where the sum is over the values of $J$ (Eq.~\ref{eq:J}) for the cusps inside the mass $\MDM$ and the angle brackets denote the average over the Monte Carlo sample of cusps obtained earlier.\footnote{Since almost all of the annihilation signal comes from prompt cusps forming by the redshift $z\sim 30$ \cite{Delos:2022bhp}, time variation in $\rhoeff$ is expected to be negligible within the redshift range relevant to $\gamma$-ray searches.}
Note that (per Eq.~\ref{eq:rhoeff_samp}) $\rhoeff$ does not depend on $M$. It also has negligible variance between regions as long as $M$ is much larger than the mass scale of individual cusps.
$\rhoeff$ is dimensionally a mass density; in this respect, it is the prompt cusps' contribution to the mass-weighted average dark matter density, since $\rhoeff= \frac{1}{\MDM}\int_\text{cusps} \rho\, \diff M$. It is a convenient quantity because the average annihilation rate $\Gamma$ per dark matter mass $\MDM$, due to prompt cusps, is
\begin{equation}\label{eq:annihilationrate}
    \langle\Gamma/\MDM\rangle_{\mathrm{cusps},0} = \frac{\sigmav}{2m_\mathrm{DM}^2}\rhoeff
\end{equation}
where $\sigmav$ is the annihilation cross section. Here we assume the dark matter is its own antiparticle and that $\sigma v$ is velocity-independent at lowest order. We insert the subscript `0' to note that this relationship does not account for the further evolution of the cusps after their formation, which we will discuss below.

\begin{figure}[tbp]
\centering
\includegraphics[width=\columnwidth]{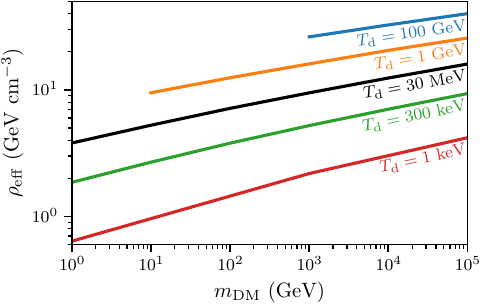}
\caption{The effective dark matter density $\rhoeff$ contributed by prompt cusps, relevant to the dark matter annihilation rate (see Eq.~\ref{eq:annihilationrate}), as a function of the particle mass $m_\mathrm{DM}$. $\rhoeff$ is the annihilation $J$ factor (proportional to the annihilation rate) per mass of dark matter, or equivalently, the mass-weighted average dark matter density. Different colors correspond to different kinetic decoupling temperatures.}
\label{fig:rhoeff}
\end{figure}

Figure~\ref{fig:rhoeff} shows how $\rhoeff$ varies with the mass $m_\mathrm{DM}$ and kinetic decoupling temperature $\Td$ of the dark matter particle. It is approximately logarithmically sensitive to both of these parameters.\footnote{Reference~\cite{Delos:2022bhp} found that $\rhoeff\simeq 0.08\left[\log\!\left(\frac{M_\mathrm{DM}}{\text{GeV}}\frac{\Td}{\text{GeV}}\right)+36\right]^5\rhoc$.} We will assume $\Td=30$~MeV for the remainder of this article, a common choice (e.g. Refs.~\cite{Green:2003un,Fermi-LAT:2015qzw}) that is relatively conservative. Models with lower $\Td$ couple more strongly to Standard Model particles, which can put them at odds with nondetection results in terrestrial experiments \cite{Cornell:2013rza}. Higher $\Td$, on the other hand, would only increase $\rhoeff$, boosting the annihilation signal further and strengthening any inferred constraints on the annihilation cross section.

Not all prompt cusps survive. Halos frequently merge together, and if they have comparable masses, it is expected that their central cusps also merge. This process effectively removes one prompt cusp while only slightly altering the other \cite{Ogiya:2016hyo,Delos:2022yhn}. Also, a substantial number of the lower initial density peaks do not even form prompt cusps because their mass accretes onto another object before they collapse. We model these effects by scaling $\rhoeff$ by a factor $\gfac< 1$. We adopt $\gfac=0.5$ based on recent simulations \cite{Delos:2022bhp}, but this quantity is uncertain at the $\sim 0.1$ ($\sim 20$ percent) level. Note that $\gfac$ propagates cleanly through our analysis, so that our constraints on $\sigmav$ can be scaled by $0.5/\gfac$ if further research refines the predicted value of $\gfac$.

In a merger between halos of highly unequal mass, the smaller object generally survives as a subhalo of the larger object. Today, almost all prompt cusps are expected to be subhalos of larger systems. However, even the smallest observationally resolved halos (e.g. those of the Milky Way's faintest satellites) are still expected to contain a very large number of prompt cusps 
($\sim 10^{10}$, perhaps). Subhalos are gradually stripped by tidal forces (e.g. \cite{Amorisco:2021hch,2022MNRAS.517.1398B,2022MNRAS.516..106D,Stucker:2022fbn}), suppressing their annihilation radiation \cite{2019PhRvD.100f3505D,2023MNRAS.518...93A}. However, this only significantly affects prompt cusps in the innermost regions of present-day halos where the dominant disruption effect is, instead, encounters with individual stars (e.g. \cite{Delos:2019tsl,Facchinetti:2022sai,Stucker:2023rjr}). In a cosmological average, these effects are generally minor.
For example, Ref.~\cite{Delos:2022bhp} used the tidal stripping model of Ref.~\cite{Stucker:2022fbn} to estimate that tidal stripping suppresses the observable annihilation signal from external galactic halos at the 10 percent level.
Similarly, Ref.~\cite{StenDelos:2019xdk} estimated that under 10 percent of dark matter resides in regions where stars are abundant, such that disruption by stellar encounters would be relevant.
Since these effects are smaller than the uncertainty in $\gfac$, we neglect them. However, if further research yields more precise modeling, the global effects of mergers, tidal stripping, and stellar encounters, as well as any small corrections to the prompt cusp density profiles (Eqs. \ref{eq:A}-\ref{eq:J}), can be absorbed at first order into $\gfac$, scaling our constraints on $\sigmav$ by $0.5/\gfac$ as discussed above.

Tidal stripping and encounters with stars are, however, highly relevant for the prompt cusp signal from our own Galactic halo as a result of our position near its center. In this context, we will denote by $\lfac(\vx)<1$ the scaling factor for the prompt cusp annihilation rate due to tidal stripping and stellar encounters, which is a function of position $\vx$.

\section{The isotropic annihilation signal}\label{sec:annihilation}

We now quantify the contribution of annihilating dark matter to the IGRB resulting from the prompt cusp distributions derived above. Let $\diff N_\gamma/\diff E$ be the average differential photon count, per energy $E$, produced by each annihilation event. We will consider $b\bar b$, $\tau^+\tau^-$, and $W^+W^-$ annihilation channels, adopting in each case the photon spectra $\diff N_\gamma/\diff E$ tabulated by Ref.~\cite{Cirelli:2010xx} (which use the results of Ref.~\cite{Ciafaloni:2010ti}).
Note that the differential photon rate emitted per mass $\MDM$ of dark matter is then  $\langle\Gamma/\MDM\rangle\diff N_\gamma/\diff E$, where $\langle\Gamma/\MDM\rangle$ is the average annihilation rate per dark matter mass.
We will neglect the contribution of dark matter not in prompt cusps. Even though only a percent-level fraction of the dark matter is in cusps, Ref.~\cite{Delos:2022bhp} used comparisons with previous halo-based predictions of the annihilation rate to show that the prompt cusps dominate the average signal.

\subsection{Galactic contribution}

A significant fraction of the dark matter contribution to the IGRB comes from our own Galactic halo (e.g. \cite{Blanco:2018esa}). Our halo has a dark matter distribution that we approximate with a spherically symmetric density profile $\rho_\mathrm{gal}(r)$. According to the above considerations, the average annihilation rate per dark matter mass at radius $r$ due to prompt cusps is
\begin{equation}\label{eq:totalrate}
    \langle\Gamma/\MDM\rangle = \frac{\sigmav}{2m_\mathrm{DM}^2}\gfac\lfac(r)\rhoeff,
\end{equation}
obtained by scaling Eq.~(\ref{eq:annihilationrate}) by the suppression factors discussed in Sec.~\ref{sec:cusps}.
As noted above, we only include the contribution of prompt cusps to the annihilation signal. While the smooth Galactic halo dominates the signal from the Galactic Center, it contributes only at the percent level beyond $\sim$$20^{\rm o}$ therefrom.

We use the \textsc{cusp-encounters} code \cite{Stucker:2023rjr} to evaluate $\lfac(r)\rhoeff$. This code adopts the description of the Galaxy's baryonic components used by Ref.~\cite{Kelley:2018pdy}, which includes axisymmetric stellar and gas disks and a stellar bulge, the parameters of which are consistent with observational constraints in Refs.~\cite{2016ARA&A..54..529B,2017MNRAS.465...76M}. The dark matter halo is taken to initially have a density profile of the Navarro-Frenk-White form \cite{Navarro:1996gj} with concentration $c\equiv R_{200c}/r_s=8.7$ and mass \mbox{$M_{200c}=10^{12}$~M$_\odot$}, but it is subjected to adiabatic contraction due to the presence of the baryons according to the prescription of Ref.~\cite{2020MNRAS.494.4291C}. We will test the impact of the choice of $c$ and $M_{200c}$. For the initial sample of prompt cusps derived in Sec.~\ref{sec:cusps}, orbits are randomly drawn from the distribution function of the Galactic halo (assuming velocity isotropy for simplicity). The prescriptions of Refs.~\cite{Stucker:2022fbn,Stucker:2023rjr} are then applied to modify the cusps' density profiles due to tidal stripping from the Galactic potential and impulsive encounters with stars along their orbits.

The photon flux that we receive from annihilation in cusps in the Galactic halo, per area and solid angle, is then
\begin{align}\label{eq:nearflux}
	\frac{\diff^2\Phi}{\diff\Omega \diff E}&=
	\frac{1}{4\pi}
	\frac{\diff N_\gamma}{\diff E}
	\int_0^\infty\diff\ell\,
	\rho(r)
    \left\langle\frac{\Gamma}{\MDM}\right\rangle
    \nonumber\\&=
	\frac{1}{4\pi}
	\frac{\diff N_\gamma}{\diff E}
    \frac{\sigmav}{2m_\mathrm{DM}^2}
	\int_0^\infty\diff\ell\,
	\rho(r)
    \lfac(r)\gfac\rhoeff,
\end{align}
where $\rho(r)$ is the density profile of the Galactic halo and \mbox{$r=\sqrt{r_0^2+\ell^2-2r\ell\cos\theta}$} is the Galactocentric radius at distance $\ell$ along the chosen line-of-sight, which is at angle $\theta$ from the Galactic Center. Here $r_0=8.2$~kpc is the distance to the Galactic Center.

\begin{figure}[tbp]
\centering
\includegraphics[width=\columnwidth]{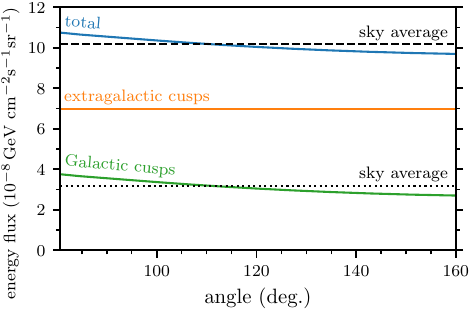}
\caption{Integrated energy flux from annihilation in prompt cusps, as a function of angle from the Galactic Center. We adopt here a 100~GeV WIMP that decouples at 30~MeV and annihilates into $b\bar b$ with a cross section of $\sigmav=10^{-26}$~cm$^3$s$^{-1}$.}
\label{fig:angular}
\end{figure}

Equation~(\ref{eq:nearflux}) is a function of both the sky angle $\theta$ and the photon energy $E$. To illustrate the angular dependence, we consider the integrated energy flux, $\int\diff E E\diff^2\Phi/\diff\Omega \diff E$. For the example model of a 100~GeV WIMP that decouples at a temperature of 30~MeV and annihilates into $b\bar b$ with cross section $\sigmav=10^{-26}$~cm$^3$s$^{-1}$, we plot the integrated flux from Galactic cusps as the green curve in Fig.~\ref{fig:angular}. In our analysis we will consider only Galactic latitudes $|b|>20$~deg. and longitudes $|l|>80$~deg., as we will discuss in Section~\ref{sec:background}. Within this region of the sky, the angle $\theta$ from the Galactic Center ranges from 80.6~deg. to 160~deg., and the annihilation signal from Galactic cusps differs from its average value (dotted line) by up to 30 percent. While the density of the Galactic halo rises strongly toward the Galactic Center, this trend is not reflected as strongly in the annihilation rate because prompt cusps nearer to the Galactic Center are suppressed more by stellar encounters and (to a lesser extent) by tidal forces, as discussed in Ref.~\cite{Stucker:2023rjr}. For dark matter models of higher mass, the prompt cusps form slightly earlier and thus have higher density, reducing their susceptibility to suppression by these effects. However, this effect is small; for a 100~TeV WIMP, the annihilation signal from Galactic cusps differs from its average value by a maximum of 38 percent.

\begin{figure}[tbp]
\centering
\includegraphics[width=\columnwidth]{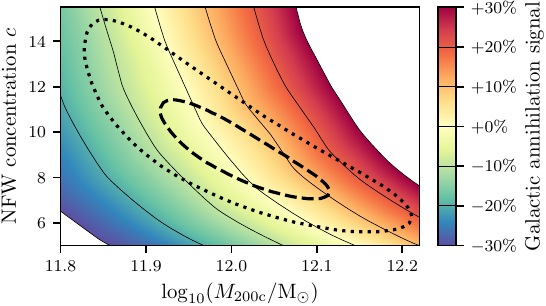}
\caption{Sensitivity of the annihilation signal from Galactic cusps to the mass $M_{200c}$ and initial NFW concentration $c$ of the Milky Way halo. We consider a 100~GeV WIMP that decouples at 30~MeV. The color scale (with contour lines every 10 percent) indicates the fractional change in the sky-averaged gamma-ray flux from Galactic cusps, compared to the fiducial parameters $c=8.7$ and $M_{200c}=10^{12}$~M$_\odot$. For comparison, the dashed and dotted curves mark the 68 and 95 percent confidence ranges, respectively, of the observational inference in Ref.~\cite{2020MNRAS.494.4291C}. Although the annihilation signal is stronger for higher $M_{200c}$ and higher $c$, its variation remains at the 10 percent level across the range of likely parameters.}
\label{fig:MWparams}
\end{figure}

The parameters of the Galactic halo are not exactly known. Figure~\ref{fig:MWparams} shows how the sky-averaged flux from annihilation in Galactic cusps changes as the halo mass $M_{200c}$ and initial NFW concentration $c$ are varied across the uncertainty range of the observational inference in Ref.~\cite{2020MNRAS.494.4291C}.
More massive and more concentrated halos yield stronger signals.
However, within the 68 percent confidence range, and even most of the 95 percent confidence range, the flux from Galactic cusps varies by under 10 percent from the fiducial value (which adopts $c=8.7$ and $M_{200c}=10^{12}$~M$_\odot$). We will neglect this uncertainty. As we will see next, Galactic cusps only source about a third of the total annihilation signal, so the total effect of the uncertainty in Galactic parameters is much smaller than 10 percent.

\subsection{Extragalactic contribution}

\begin{figure}[tbp]
\centering
\includegraphics[width=\columnwidth]{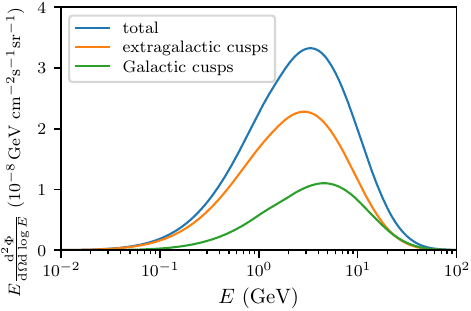}
\caption{Differential energy flux from annihilation in prompt cusps, averaged over latitudes $|b|>20$~deg. and longitudes $|l|>80$~deg. As in Fig.~\ref{fig:angular}, we assume a 100~GeV WIMP that decouples at a temperature of 30~MeV and annihilates into $b\bar b$ with a cross section of $\sigmav=10^{-26}$~cm$^3$s$^{-1}$.}
\label{fig:spectrum}
\end{figure}

The average annihilation rate per dark matter mass in the extragalactic field is
\begin{equation}
    \langle\Gamma/\MDM\rangle = \frac{\sigmav}{2m_\mathrm{DM}^2}\gfac\rhoeff.
\end{equation}
The photon flux from annihilation in extragalactic cusps is
\begin{align}\label{eq:farflux}
	\frac{\diff^2\Phi}{\diff\Omega \diff E}
    &=
	\frac{1}{4\pi}\left\langle\frac{\Gamma}{\MDM}\right\rangle\rhoc
	\int_0^\infty\frac{\diff z}{H(z)/c}
	\left.\frac{\diff N_\gamma}{\diff E^\prime}\right|_{E^\prime = E(1+z)}
    \nonumber\\&=
	\frac{1}{4\pi}\frac{\sigmav}{2m_\mathrm{DM}^2}\gfac\rhoeff\rhoc
	\!\!\int_0^\infty\!\!\!\!\!\frac{\diff z}{H(z)/c}
	\left.\frac{\diff N_\gamma}{\diff E^\prime}\right|_{E^\prime = E(1+z)}\!\!.
\end{align}
As we discussed in Sec.~\ref{sec:rhoeff}, tidal stripping and stellar encounters (incorporated via $\lfac$ in Eq.~\ref{eq:nearflux}) are small effects in the cosmological average, so we do not explicitly include them here.
Due to the cosmological redshift, extragalactic cusps contribute with a different energy spectrum from Galactic cusps. This difference is illustrated in Fig.~\ref{fig:spectrum} for the same example 100~GeV WIMP model. Extragalactic cusps (orange curve) contribute about twice the energy flux of Galactic cusps (green curve), and their contribution is skewed toward slightly lower energies.

The integrated energy flux from extragalactic cusps is shown in Fig.~\ref{fig:angular} (orange line) for the same model. It is evident again that extragalactic cusps contribute at about twice the level of Galactic cusps. We also show the total cusp signal (blue curve), summing Galactic and extragalactic contributions. Within the region of the sky under consideration, at Galactic latitudes $|b|>20$~deg. and longitudes $|l|>80$~deg., the total signal differs from its average value (dashed line) by well under 10 percent. Such a small variation would be indistinguishable from an isotropic signal, and thus for the remainder of the analysis, we assume that the dark matter signal is isotropic and consider only its average value.

\section{Isolating the observed isotropic background flux}\label{sec:background}

With the predicted dark matter contribution to the IGRB established, we now begin our analysis of the observed IGRB using 14 years of Pass 8 Fermi-LAT data.
We first separate the diffuse isotropic emission from anisotropic foreground emission components using a morphological analysis. The $\gamma$-ray foreground includes contributions from Galactic and extragalactic point sources, as well as diffuse Galactic $\gamma$-ray emission. To separate these components, we use a model very similar to ``Model A’’ in Ref.~\cite{Widmark2023}, with a few modifications. In particular, we use different angular cuts: $|l|>80^\circ$ and $|b|>20^\circ$, which excludes slightly more of the Galactic disk and Galactic Center. The excluded regions are more strongly affected by foregrounds and hence provide lower signal-to-noise when analyzing the IGRB.

The diffuse Galactic emission comes from a combination of the inverse Compton scattering (ICS) of starlight by relativistic electrons, as well as the bremsstrahlung and hadronic interactions of relativistic electrons and protons, respectively, with Galactic gas. The morphology of this emission closely traces the Galactic gas density, which is traced in turn by a combination of several key gas tracers, including:

\begin{itemize}
\item {\bf CO --- }We use the radio observations of carbon monoxide (CO) from the Planck survey \cite{2014A&A...571A..13P} as a tracer of cold molecular hydrogen gas. The component of CO-traced H$_2$ has a small scale height, and thus, due to our cut on Galactic latitude, gives only a minuscule contribution.
\item {\bf Dust --- }We use a three-dimensional dust map based on \emph{Gaia} data \citep{2019A&A...625A.135L} to trace the cold neutral medium (CNM).
\item {\bf 21 cm --- }We use the two-dimensional full-sky 21 cm map called HI4PI \cite{2016A&A...594A.116H}, as a tracer of atomic hydrogen.
\end{itemize}
We refer to Ref.~\cite{Widmark2023} for a more detailed account of these $\gamma$-ray components.

We infer the $\gamma$-ray source normalizations in a Bayesian framework, using Hamiltonian Monte Carlo, where each energy bin is fitted separately. We use strictly positive wide flat box priors for all $\gamma$-ray components. Furthermore, we perform a jackknife binning of the sky area to quantify systematic uncertainties. We divide the sky into ten jackknife regions, which all cover half of the non-masked sky. These regions are defined as follows.
\begin{enumerate}
    \item North: $b \geq 0^\circ$,
    \item South: $b < 0^\circ$,
    \item East: $l < 180^\circ$,
    \item West: $l \geq 180^\circ$,
    \item Closer to Galactic Center (G.C.): $\text{abs}(l-180^\circ) > 50^\circ$,
    \item Towards anti-G.C.: $\text{abs}(l-180^\circ) \leq 50^\circ$,
    \item Joined south-east and north-west: ($b\geq 0^\circ$ and $l < 180^\circ$) or ($b\le 0^\circ$ and $l \geq 180^\circ$),
    \item Joined south-west and north-east: ($b\geq 0^\circ$ and $l \geq 180^\circ$) or ($b\le 0^\circ$ and $l < 180^\circ$),
    \item Joined south-G.C. and north-anti-G.C.: ($b\geq 0^\circ$ and $\text{abs}(l-180^\circ) > 50^\circ$) or ($b\le 0^\circ$ and $\text{abs}(l-180^\circ) \leq 50^\circ$),
    \item Joined south-anti-G.C. and north-G.C.: ($b\geq 0^\circ$ and $\text{abs}(l-180^\circ) \leq 50^\circ$) or ($b\le 0^\circ$ and $\text{abs}(l-180^\circ) > 50^\circ$),
\end{enumerate}
For each jackknife region and energy bin, we sample the posterior density using Markov chain Monte Carlo (MCMC). After a thorough burn-in phase, where the mode is located and the step-size is tuned, the MCMC is run for 80 000 steps. The variance of the full chain is roughly equal to the variance between realizations separated by 20 steps. We thin the chain by this factor, giving a total of 4000 MCMC realizations. The uncertainty given by the difference between jackknife regions is dominant and significantly larger than the statistical uncertainty associated with the individual MCMC chains.

\begin{figure}[tbp]
\centering
\includegraphics[width=0.48\textwidth]{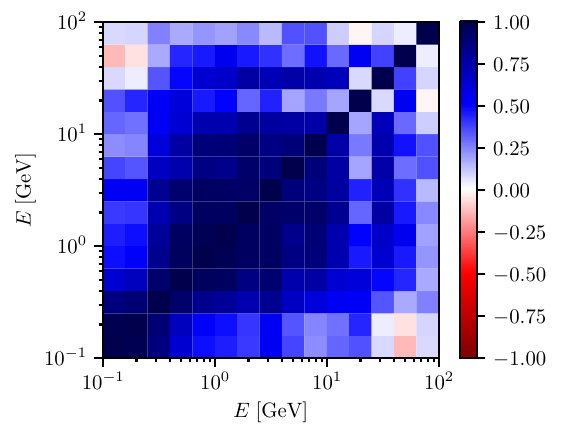}
\caption{Correlation coefficients for the 15 energy bins of the isotropic component. The diagonal is valued unity by definition. Otherwise nearby energy bins typically have strong positive correlations.}
\label{fig:Ecorrs}
\end{figure}

The different jackknife regions not only allow us to quantify systematic uncertainties, but they also enable us to estimate possible correlations between the respective energy bins. If the fitted IGRB in different energy bins has the same upward or downward fluctuations in the same jackknife regions, that implies that they are subject to a similar fitting error; in other words, there is a positive correlation between those energy bins. We calculate the fitted IGRB correlation matrix over the different energy bins, by taking the correlation factor of the respective jackknife region's MCMC realizations.\footnote{
In exact terms, the correlation factor between two energy bins, labeled by indices $i$ and $j$, is calculated according to:
\begin{equation}
    \text{Corr. factor}_{(i, j)} =  \frac{\sum_{k=1}^{10} \sum_{n=1}^{4000} (x_{i,k,n}-\bar{x}_i) (x_{j,k,n}-\bar{x}_j)}{4 \times 10^4 \, \text{Std.}(x_i) \, \text{Std.}(x_j)},
\end{equation}
where $k$ denotes the jackknife region, $n$ denotes the MCMC realization index, and $x_{i,k,n}$ is an MCMC realization of the IGRB normalization factor. The quantities $\bar{x}_i$ and $\text{Std.}(x_i)$ represent the mean and standard deviation of the MCMC chains of the $i$th energy bin, all jackknife regions included. The numerical factor in the denominator is equal to the product of the number of jackknife regions and number of MCMC iterations.}
The correlation matrix is visible in Fig.~\ref{fig:Ecorrs}. Although this approach gives a handle on correlations, it is not complete; in principle, there could also be systematic correlations between energy bins that are not dependent on sky angle, which we would not be sensitive to with this procedure.

The black points in the upper panel of Fig.~\ref{fig:IGRB_fit_results} show the isotropic energy flux that we obtain by this procedure. We note that this is not exactly the IGRB, because it includes likely contamination by misidentified cosmic rays, which we will discuss and account for in Sec.~\ref{sec:limits}.

\section{Astrophysical contributions to the IGRB}
\label{sec:IGRB Model}

\begin{figure}[t]
\centering
\includegraphics[width=0.48\textwidth]{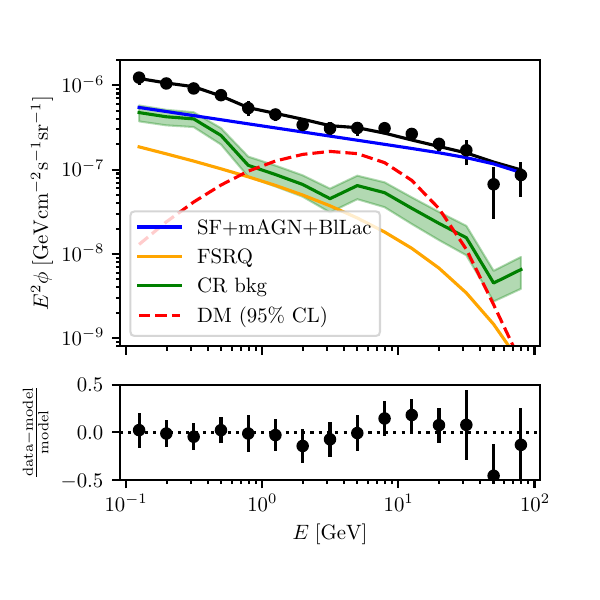}
\caption{The diffuse isotropic emission (black points) and the results of our fit to it (black curve). This emission includes contributions from cosmic rays that have been misidentified as $\gamma$ rays by the Fermi-LAT instrument, $\gamma$-ray contributions from extragalactic star-forming activity, blazar and non-blazar AGN source classes, as well as potential contributions from dark matter annihilation. We find that our combined model, including astrophysical $\gamma$-ray emission (blue and orange curves) and misidentified cosmic rays (green curve), fits the observed data without any evidence for an excess from dark matter annihilation. Here we show the 95 percent confidence limit on the emission from dark matter with a mass of 100 GeV annihilating into $b$ quarks (dashed curve).
}
\label{fig:IGRB_fit_results}
\end{figure}

The majority of the IGRB is known to originate from astrophysical sources (e.g.~\cite{Blanco:2018esa,Blanco:2021icw}).
In order to model these contributions, we consider spectral components from blazar and non-blazar sources. Our non-blazar component includes extragalactic star-forming activity (SF) and misaligned AGN (mAGN). Our blazar component includes BL Lacertae objects (BL Lacs) and flat spectrum radio quasars (FSRQs). We adopt the IGRB model in Ref.~\cite{Blanco:2021icw} and allow the relative contribution of each source class to vary according to the log-normal distributions found therein. This model accounts for the potential IGRB contribution from any galaxy that hosts a non-blazar supermassive black hole since it allows for misaligned AGN gamma-ray emission to be considered independently from the gamma-ray emission originating from its star-forming activity. 

This model also accounts for the electromagnetic cascades that originate from $\gamma$-rays above about 100 GeV as they generate high-energy electrons when scattering with the extragalactic background light (EBL+CMB)~\cite{Dominguez:2010bv,Blanco:2018bbf}. In turn, these electrons produce gamma rays after inverse-Compton scattering with the same radiation fields. It is important to account for these effects since the spectral shape of any sufficiently energetic source class is significantly altered by these cascades, especially above 100 GeV. While the blazar model that we adopt does not account for cascade production, we note that (1) the contribution from unresolved blazars cannot be significantly enhanced by their cascades~\cite{Blanco:2023kfa} and (2) the blazar contribution is expected to be subdominant to SF and mAGN at all relevant gamma-ray energies~\cite{Blanco:2021icw}. 

Since there is a considerable degeneracy in the spectral shape of the SF, mAGN, and BL Lac contributions, their spectral shape is considered to be the same while the normalization is fit using their joint log-normal distribution as a prior. We note that the summed log-normal distribution can be directly calculated by using the full covariance matrix for the individual uncertainties calculated in Table II of Ref.~\cite{Blanco:2021icw}. Because the spectrum of the FSRQ contribution differs from the remaining astrophysical components, the FSRQ source class is considered independently. As shown in Fig.~\ref{fig:IGRB_fit_results}, our model can account for the entirety of the measured isotropic gamma rays, allowing us to constrain the addition of a potential dark matter signal.

\section{Dark matter limits}\label{sec:limits}

\begin{figure*}[t]
\centering
\includegraphics[trim={0.5cm 0.2cm 0.5cm 0.0cm},clip, width=0.25\textwidth]{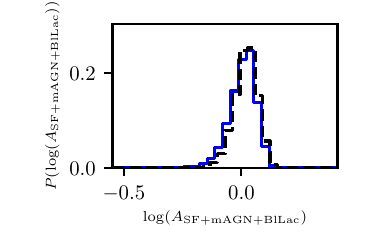}%
\includegraphics[trim={0.5cm 0.2cm 0.5cm 0.0cm},clip, width=0.25\textwidth]{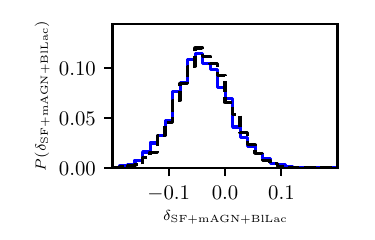}%
\includegraphics[trim={0.5cm 0.2cm 0.5cm 0.0cm},clip, width=0.25\textwidth]{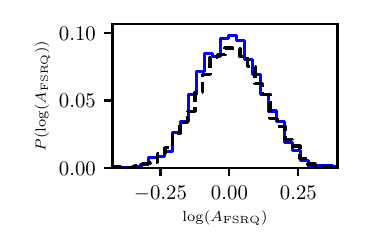}%
\includegraphics[trim={0.5cm 0.2cm 0.5cm 0.0cm},clip, width=0.25\textwidth]{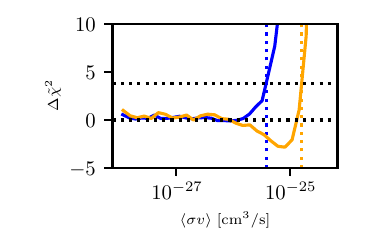}
\caption{
The three panels on the left display the marginalized posterior distribution for the fit parameters of the astrophysical model. Dashed black lines show the fit without dark matter while the solid blue lines show the result of the fit for a fixed dark matter mass of 100 GeV and annihilation into $b$ quarks. The rightmost panel displays the marginalized $\chi^2$ as a function of the dark matter annihilation cross section. The blue (yellow) line corresponds to a dark matter mass of 100 (300) GeV. The vertical dotted lines indicate the 95\% upper limit on the dark matter annihilation cross section. }
\label{fig:IGRB_fit_posteriors}
\end{figure*}

Dark matter annihilation and astrophysical sources would both contribute to the diffuse isotropic emission that we obtained in Sec.~\ref{sec:background}.
To obtain a limit on the dark matter annihilation cross section, we now employ a combination of spectral information and multiwavelength constraints to differentiate the contributions from each component. 
In particular, we utilize a template fit approach that incorporates the astrophysical sources discussed in Sec.~\ref{sec:IGRB Model}, the contamination from cosmic ray (CR) backgrounds, and the Galactic and extragalactic dark matter signal outlined in Sec.~\ref{sec:annihilation}.

To account for uncertainties in the model of astrophysical sources and the CR background contamination, we introduce nuisance parameters. Subsequently, we perform a template fit to the measurement of the isotropic flux using four templates: SF+mAGN+BlLac, FSRQ, CR background contamination, and a dark matter model. The astrophysical contribution of SF, mAGN, and BlLac have similar spectral shapes and exhibit significant degeneracy in normalization, whereas the sum of their contributions is more tightly constrained. Hence, we opt to consider a single template that encompasses all three populations. In our default simulations, the CR background is set to the results of Ref.~\cite{Fermi-LAT:2014ryh} and represents instrumental misidentification of charged CR events as a $\gamma$-ray flux. We note that Ref.~\cite{Fermi-LAT:2014ryh} was calculated using Pass 7 data, and we discuss in detail systematic tests of uncertainties in the CR background below. 

The free parameters involved in the fit are:
\begin{itemize}
    \item The normalization, $A_{\rm SF+mAGN+BlLac}$, of the astrophysical contribution from SF+mAGN+BlLac with respect to the model from Ref.~\cite{Blanco:2021icw}. We choose a Gaussian prior in $\log_{10} A_{\rm SF+mAGN+BlLac}$ centered at $\mu=0$ with width $\sigma$ determined by the uncertainty from Ref.~\cite{Blanco:2021icw}.  
    \item The deviation of the spectral slope, $\delta_{\rm SF+mAGN+BlLac}$, of the SF+mAGN+BlLac contribution with respect to the reference model. We apply a flat prior for between $-0.2$ and $0.2$.  
    \item The normalization, $A_{\rm FSRQ}$, of the astrophysical FSRQ contribution. The prior is again Gaussian in $\log_{10} A_{\rm FSRQ}$ with $\mu=0$ and $\sigma$ taken from Ref.~\cite{Blanco:2021icw}.
    \item The dark matter annihilation cross section $\sigmav$. We employ a flat prior in $\log_{10}(\sigmav / \rm cm^3s^{-1} )$ between $-28$ and $-24$.
\end{itemize}
We perform 62 fits, one without dark matter (i.e. $\sigmav$ fixed to 0) and the remaining 61 fits using discrete values of the dark matter mass approximately logarithmically spaced between 5 GeV and 100 TeV.
For each fit we minimize the negative log-likelihood defined as
\begin{equation}
    \label{eq:likelihood}
    -2\log \mathcal{L} = \sum\limits_{i,j} (\phi^{(m)}_i - \phi^{(d)}_i) [\mathbf{C}^{-1}]_{ij}(\phi^{(m)}_j - \phi^{(d)}_j),
\end{equation}
where $\mathbf{C}^{-1}$ is the matrix inverse of
\begin{equation}
    C_{ij}=V_{ij} + \delta_{ij} (\sigma^{(\rm CR,bkg)}_i)^2.
\end{equation}
Here $\phi^{(d)}$ is the measured isotropic flux with the covariance matrix $V$, $\phi^{(m)}$ is the sum of the model templates, and $\sigma^{(\rm CR,bkg)}$ is the uncertainty of the CR background contamination extracted from Ref.~\cite{Fermi-LAT:2014ryh}. 

For each fit, we perform a comprehensive MultiNest parameter scan to obtain the full posterior distribution. First we analyze the fit without dark matter. By utilizing solely the astrophysical source and CR contamination model, we achieve a satisfactory fit to the data. Moreover, the posterior distributions of the nuisance parameters, introduced for the astrophysical source populations, align well with the reference model normalizations, and the deviation of the slope is consistent with zero. The marginalized posteriors for each parameter are visualized by the dashed black lines in Fig.~\ref{fig:IGRB_fit_posteriors}.
The $\chi^2$ value amounts to 17.5 for 15 data points, allowing us to proceed with deriving dark matter limits.

\begin{figure}[t]
\centering
\includegraphics[width=0.48\textwidth]{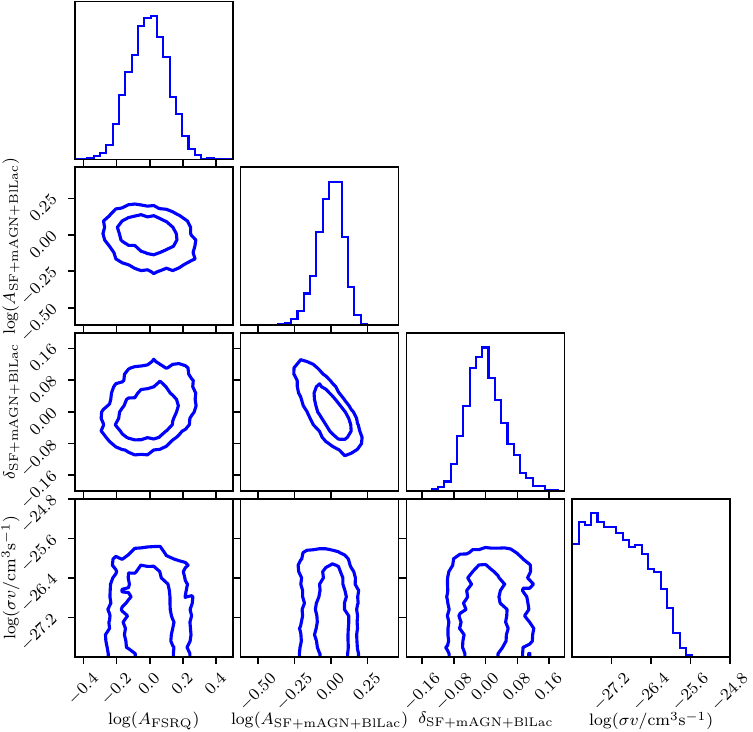}
\caption{
Posterior distribution of the three parameters of the astrophysical model and the dark matter annihilation cross section, for a dark matter mass of 100~GeV. The inner and outer contours enclose 68 and 95 percent of the distribution, respectively.
}
\label{fig:IGRB_fit_posteriors_triangle}
\end{figure}

Considering an example involving dark matter, the blue lines in Fig.~\ref{fig:IGRB_fit_posteriors} show the posteriors for $m_{\rm DM} = 100$~GeV and annihilation into $b$ quarks. In this section, we utilize annihilation into $b$ quarks as an example to investigate systematic uncertainties. However, we will present the limits for annihilation into $\tau$ leptons and $W$ gauge bosons in the concluding section. At 100 GeV there is no preference for dark matter, as can be seen from multiple angles. First, the $\chi^2$ of the fit does not improve. Second, the posterior distribution of the parameters describing the astrophysical templates are not affected by the additional dark matter template. Third, the posterior distribution in $\sigmav$ is flat at low cross sections. 
We also show the posterior distribution in more detail in Fig.~\ref{fig:IGRB_fit_posteriors_triangle}. Notice in particular that there is no degeneracy between the dark matter cross section and astrophysical parameters.

The dark matter limit is determined from the difference of the marginalized $\chi^2$ between the cases with and without dark matter annihilation, which is given by 
\begin{eqnarray}
    \Delta \tilde{\chi}^2(\sigmav) 
    = -2\log\Big( \frac{\mathcal{L}(\sigmav)}{\mathcal{L}(\sigmav=0)} \Big)\,.
\end{eqnarray}
Here $\mathcal{L}(\sigmav)$ is the likelihood defined in Eq.~\eqref{eq:likelihood} that is marginalized over the vector of astrophysical parameters (denoted $\theta$), namely: 
\begin{eqnarray}
\mathcal{L}(\sigmav) = \int \diff \theta \, \pi(\theta) \mathcal{L}(\theta,\sigmav) \, , 
\end{eqnarray}
where $\pi(\theta)$ is the prior distribution in $\theta$. Finally, the dark matter limit at 95\% confidence level is obtained at $\Delta \tilde{\chi}^2(\sigmav)=3.84$.\footnote{Another possibility would be to identify the $\sigmav$ below which 95\% of the posterior distribution lies. Although this approach yields slightly stronger dark matter limits for our priors, we do not use it because the result is explicitly sensitive to the lower limit of the prior on $\sigmav$.}
Technically, $\mathcal{L}(\sigmav)$ is obtained from the MultiNest scans with dark matter annihilation, while $\mathcal{L}(\sigmav=0)$ is taken from the scan without it.%
\footnote{
We note that the marginalized likelihood is related to the marginalized posterior, $P(\sigmav)$, by $\mathcal{L}(\sigmav) = P(\sigmav) Z/\pi(\sigmav)$, where $Z$ is the evidence and $\pi(\sigmav)$ is the prior in $\sigmav$. Furthermore, $\mathcal{L}(\sigmav=0)$ is equal to the evidence of the fit without dark matter.}
The rightmost panel in Fig.~\ref{fig:IGRB_fit_posteriors} shows the marginalized $\chi^2$ and the dark matter limit for the example dark matter masses of 100 and 300 GeV.

In Fig.~\ref{fig:IGRB_fit_results} we compare our model with the isotropic flux data (black points). We show fits for the astrophysical templates that correspond to the median parameters at the peak of the marginalized posteriors, while the dark matter template is shown for $\sigmav$ at the 95\% limit. 
The black line shows the sum of the astrophysical templates (i.e. without dark matter). The residuals, defined as (data-model)/model, in the lower panel are also evaluated for a model that does not include the dark matter template. 

\begin{figure}[t]
\centering
\includegraphics[width=0.48\textwidth]{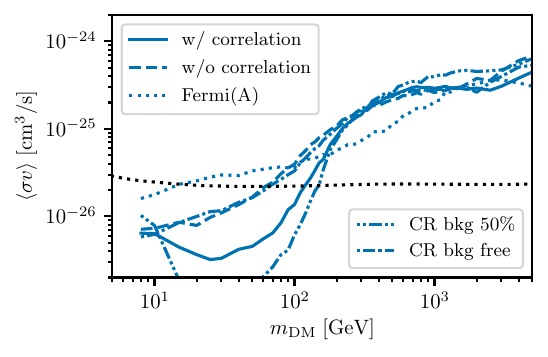}
\caption{Upper limits on dark matter annihilation into $b$ quarks. The solid curve is our fiducial result, while the other blue curves test the sensitivity of the result to different assumptions (see the text for more detail). The dotted black line shows the annihilation cross section expected for a thermal WIMP~\cite{Steigman:2012nb}.}
\label{fig:DM_limits_sys}
\end{figure}

\begin{table*}[t]
\caption{Results of the MultiNest scans, comparing fits with and without the inclusion of dark matter. The fits with dark matter (DM) are based on two additional free parameters: the dark matter mass and the annihilation cross section. The provided values include the best-fit $\chi^2$ for each case, along with the corresponding log-evidences. We also provide the dark matter mass and the annihilation cross section for the best fit. While the inclusion of dark matter shows slight improvements in the fits, these enhancements are not statistically significant (see text for details).
}
\label{tab:fit_results}
\begin{tabular}{ l c cccccccccc }
 \hline \hline
 Fit & $\quad$ &  $\chi^2$ w/o DM & $\quad$ & $\chi^2$ w/ DM & $\quad$ & $\ln\mathcal{Z}$ w/o DM & $\quad$ &  $\ln\mathcal{Z}$ w/ DM  &$\quad$ & $m_{\rm DM}\,\mathrm{[GeV]}$ & $\sigmav\,\mathrm{[10^{26} cm^3/s]}$   \\
\hline 
$b\bar{b}$                 & & 17.5 & & 7.7 & &  $-11.4 \pm  0.040$  & & $-10.6  \pm  0.060$ & & $ 454 $ & $ 15    $  \\
$\tau^+\tau^-$             & & 17.5 & & 5.4 & &  $-11.4 \pm  0.040$  & & $-10.0  \pm  0.071$ & & $  49 $ & $ 2.5   $  \\
$W^+W^-$                   & & 17.5 & & 8.8 & &  $-11.4 \pm  0.040$  & & $-10.9  \pm  0.050$ & & $ 305 $ & $ 15    $  \\
$b\bar{b}$ w/o correlation & &  4.6 & & 1.9 & &  $ -5.3 \pm  0.044$  & & $ -5.7  \pm  0.051$ & & $ 345 $ & $ 11    $  \\
$b\bar{b}$ CR bkg. 50\%    & &   43 & &  20 & &  $-19.5 \pm  0.081$  & & $ -24.8 \pm  0.044$ & & $ 474 $ & $ 17    $  \\
$b\bar{b}$ CR bkg. free    & & 11.6 & & 6.0 & &  $-10.1 \pm  0.052$  & & $-10.1  \pm  0.062$ & & $ 402 $ & $ 13    $  \\
$b\bar{b}$ Fermi           & &  4.2 & & 3.5 & &  $ -5.3 \pm  0.046$  & & $ -6.0  \pm  0.053$ & & $ 364 $ & $ 0.016 $  \\
 \hline \hline
\end{tabular}
\end{table*}

The cross-section limit for dark matter masses ranging from 5 GeV to 5 TeV is represented by the solid blue line in Fig.~\ref{fig:DM_limits_sys} As expected, stronger limits are found for smaller dark matter masses. The precise increase of the upper limit results from the interplay between the decreasing number density of dark matter particles (scaling with $m_{\rm DM}^{-1}$) and the decrease of the astrophysical background flux at higher energies. At 10 GeV, we obtain a limit of about $6\times10^{-27}\ \mathrm{cm^3/s}$ for dark matter annihilation to a $b\bar{b}$ final state, which then rises to approximately $4\times10^{-25}\ \mathrm{cm^3/s}$ at 1 TeV.

We note that around 500 GeV the dark matter limit appears relatively weak. This is due to a slight preference for the dark matter template. However, this preference is not statistically significant. To assess this, we conduct an additional fit where the dark matter mass is treated as a free parameter. The resulting $\chi^2$ difference between the fit with and without dark matter is 9.8, corresponding to a local significance of less than 3$\sigma$ for two additional fit parameters employing the frequentist interpretation. Furthermore, the Bayes factor, which is the ratio of the fit with and without dark matter, is 2.2, leads to the same conclusion.
The best $\chi^2$ and evidences of all fits are summarized in Tab.~\ref{tab:fit_results}.
The significance of the dark matter preference is not robust against the systematic uncertainties (see discussion below) and, therefore, should not be taken seriously.

While our default approach involves incorporating correlated uncertainties for the isotropic flux measurement, we also evaluate our constraints in models where we do not include correlations between different energy bins. This approach is more conservative since it allows for greater flexibility in both the astrophysical model and dark matter, as the shape is less constrained by the correlation. This is evident from the minimal $\chi^2$ values, which decrease to 4.6 in the fit without dark matter and 1.9 in the fit with dark matter. Consequently, the $\chi^2$ per degree of freedom is significantly lower than one, providing further evidence that our estimation of the covariance matrix is reasonable.
The dark matter limit obtained for this case is comparable to the default scenario. Only for $M_\mathrm{DM}$ ranging from 20 to 80 GeV does this approach provides a slightly weaker limit by about a factor of 2 as shown in Fig.~\ref{fig:DM_limits_sys}.

Another systematic uncertainty concerns the CR background contamination taken from Ref.~\cite{Fermi-LAT:2014ryh}. As detailed above, in the benchmark analysis we assume that the CR contamination rate in our analysis is the same. While this is a reasonable assumption, we note that there are some differences between our analysis and Ref.~\cite{Fermi-LAT:2014ryh} that might impact the CR contamination. Ref.~\cite{Fermi-LAT:2014ryh} uses specific additional cuts to the \texttt{Pass 7} data with \texttt{ULTRACLEAN} veto in their analysis to decrease the CR contamination systematically, separately at low and high energies. We employ the updated \texttt{Pass 8} data, which are supposed to feature a better background rejections but on the other hand we do not use additional cuts to suppress the contamination. 

To address our incomplete knowledge of the CR contamination we performed two additional tests. First, we fixed the CR background template to 50\% of our benchmark and, second, we allowed the normalization of the CR background (but not its spectrum) to vary as a free parameter in the fits.
When the CR background template is fixed at 50\%, we do not achieve a satisfactory fit to the isotropic flux using only the astrophysical model. The best-fit $\chi^2$ value increases to 43, indicating a poor fit. Furthermore, the nuisance parameter $\delta_{\rm SF+mAGN+BlLac}$ converges to the lower boundary at $-0.2$. Consequently, we obtain a stronger dark matter limit, particularly at a few tens of GeV. However, this limit should be treated with caution. Conversely, when the normalization of the CR contamination template is allowed to vary as a free parameter, it converges to approximately 175\%. In this scenario, the limit becomes more conservative, with the most significant impact observed once again at a few tens of GeV. We note that while neither of these techniques are optimal, the Fermi-LAT data on cosmic-ray contamination for the Pass 8 dataset has not been publicly released, making it difficult to self-consistently analyze the rate of cosmic-ray contamination. The approach shown here is effective in bracketing the possible levels of cosmic-ray contamination. 

The summary presented in Table \ref{tab:fit_results} shows that the preference for a dark matter signature is not robust against different systematic assumptions. While the local significance is measured at 2.6$\sigma$ for our benchmark case, it diminishes to 1.9$\sigma$ when the cosmic ray (CR) background contamination is allowed to vary in the fit, and further drops to 1.1$\sigma$ in the case without correlations.
Only in the scenario with CR background contamination fixed at 50\% the significance rises to approximately 4.4$\sigma$. However, this particular case fails to provide a satisfactory fit to the overall data.
This discussion highlights the critical importance of accurately estimating the CR background contamination for future dark matter searches. The importance of the CR background can also be seen from Fig.~\ref{fig:IGRB_fit_results}: The \emph{bump-like} spectral shape of the CR background resembles that of a dark matter signature leading to a degeneracy with a potential dark matter signal. 

As a final cross-check, we directly use the official IGRB measurement conducted by Fermi \cite{Fermi-LAT:2014ryh} in 2014 to derive the dark matter limit. In this particular case, the CR background contamination is already subtracted by the collaboration, thus eliminating its influence. Additionally, it is important to note that we cannot consider correlations in this case, since they are not provided by the collaboration.
The limit obtained using this dataset is comparable, but it exhibits fewer discernible features and a smaller preference for the dark matter template. This can be attributed to the larger uncertainties associated with the IGRB measurement itself.

\begin{figure}[t]
\centering
\includegraphics[width=0.48\textwidth]{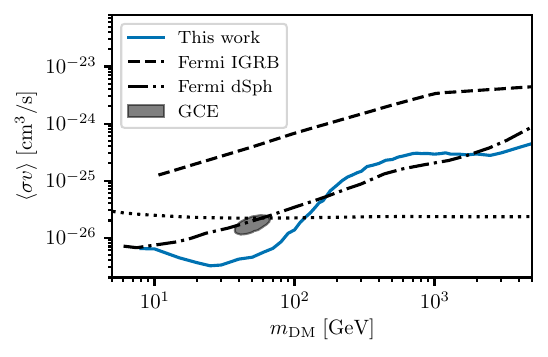}
\caption{Comparison with other results. This figure plots the upper limit on $\sigmav$ as a function of the dark matter particle mass $m_\mathrm{DM}$ for the case of annihilation into $b$ quarks. The solid curve shows our constraint using prompt cusps, while the dashed curve shows the Fermi Collaboration's constraint using a halo model \cite{Fermi-LAT:2015qzw}; both of these constraints are based on measurements of the IGRB. The dot-dashed curve shows the Fermi Collaboration's limit from dwarf spheroidal galaxies \cite{Fermi-LAT:2016uux}.
The shaded region is compatible with an annihilation origin for the Galactic Center $\gamma$-ray excess~\cite{Calore:2014xka}.
}
\label{fig:DM_limits_compare}
\end{figure}

\section{Discussion and Conclusions}

We have examined the consequences of highly concentrated prompt cusps of dark matter that form at the onset of nonlinear structure formation and survive throughout the bulk of present-day dark matter halos. Due to the contribution of these systems, we find that strong WIMP annihilation constraints come from the isotropic $\gamma$-ray background. These constraints can be stronger than those from targeted studies of regions with high central density, such as dwarf spheroidal galaxies. 

Utilizing 14 years of Fermi-LAT data, we produced a state-of-the-art model for the isotropic $\gamma$-ray background. We fit this model with a combination of astrophysical components produced by star-forming galaxies, misaligned AGN, blazars, and residual cosmic-ray contamination. We found no evidence for a remaining excess that might be due to dark matter annihilation, and so we are able to set strong limits on the dark matter annihilation cross-section. These limits are shown in Fig.~\ref{fig:constraints} for dark matter annihilating into $b$ quarks, $\tau$ leptons, or $W$ bosons.

In Figure~\ref{fig:DM_limits_compare}, we compare the results of our analysis with the Fermi collaboration's bound on the dark matter annihilation cross section \cite{Fermi-LAT:2015qzw} using the IGRB measurement from Pass 7 Fermi-LAT data \cite{Fermi-LAT:2014ryh}. This result used substructure models that did not account for prompt cusps.
Our IGRB-based limits are stronger by more than an order of magnitude throughout most of the mass range, allowing us to exclude with 95\% confidence thermally annihilating WIMP models with particle masses below 120~GeV (for a $b\bar{b}$ final state). As noted by Ref.~\cite{Delos:2022bhp}, the annihilation rate in prompt cusps exceeds the annihilation rate assumed by Ref.~\cite{Fermi-LAT:2015qzw} by about a factor of 5. Thus, prompt cusps are responsible for the majority of the improvement, while the additional years of data and the IGRB astrophysical model that we adopt (Sec.~\ref{sec:IGRB Model}) are likely responsible for the rest.

We additionally compare our results to Fermi-LAT limits on annihilation in dwarf spheroidal galaxies from Ref.~\cite{Fermi-LAT:2016uux} (which differ only marginally from a recent update \cite{McDaniel:2023bju}) as well as dark matter fits to the Galactic Center $\gamma$-ray excess~\cite{Calore:2014xka}. The existence of prompt cusps is not expected to significantly change these estimates, because these searches target regions that are of such high average density that the smooth dark matter component should dominate the annihilation rate. We find that our IGRB constraints are stronger than the dwarf constraints throughout most of the parameter space, indicating that if present treatments of prompt cusps are accurate, then Fermi-LAT observations of the IGRB provide among the strongest indirect constraints on velocity-independent annihilation of $\mathcal{O}(100)$~GeV dark matter.\footnote{$\gamma$ rays from small nearby systems can produce potentially stronger constraints, depending on assumptions about their unclear astrophysical nature (e.g.~\cite{Chan:2022amt,Crnogorcevic:2023ijs}). Alternative messengers, such as cosmic rays (e.g.~\cite{Balan:2023lwg,DelaTorreLuque:2024ozf}) and radio waves (e.g.~\cite{Regis:2021glv,Guo:2022rqq}), may also produce strong limits.} Additionally, our constraints are in tension with the cross sections necessary for dark matter to produce the Galactic Center excess. 

Finally, we note that the annihilation limits that we set in this work depend strongly on recent advances in our understanding of the astrophysical composition of the IGRB. Future $\gamma$ ray and multiwavelength studies capable of decreasing the systematic uncertainty in our understanding of extragalactic star formation activity and mAGN contributions to the IGRB thus offer the potential to unlock extremely sensitive searches for dark matter annihilation with current and future $\gamma$-ray observations.\\

\section*{Acknowledgements}
The work of C.B.~was supported in part by NASA through the NASA Hubble Fellowship Program grant HST-HF2-51451.001-A awarded by the Space Telescope Science Institute, which is operated by the Association of Universities for Research in Astronomy, Inc., for NASA, under contract NAS5-26555.
M.K. and T.L. acknowledge support by the Swedish Research Council under contracts 2019-05135 and 2022-04283, and along with C.B., also acknowledge  support from the European Research Council under grant 742104.
T.L. is also supported by the Swedish National Space Agency under contract 117/19.
A.W. acknowledges support from the Carlsberg Foundation via a Semper Ardens grant (CF15-0384).
This project used computing resources from the Swedish National Infrastructure for Computing (SNIC) and National Academic Infrastructure for Supercomputing in Sweden (NAISS) under project Nos. 2021/3-42, 2021/6-326, 2021-1-24 and 2022/3-27, partially funded by the Swedish Research Council through grant no. 2018-05973.
\clearpage

\bibliography{main}

\end{document}